\documentclass[10pt]{iopart}
\usepackage{graphicx}

\begin{document}
\paper{Local correlations, non-local screening, multiplets, and band formation in NiO}
\author{T Haupricht$^1$, J Weinen$^{1,2}$, A Tanaka$^3$, R Gierth$^1$, S G Altendorf$^1$, Y-Y Chin$^{1,2,4}$,
T Willers$^1$, J Gegner$^1$, H Fujiwara$^1$\footnote{Present address: Graduate School of Engineering Science, Osaka University, 1-3 Machikaneyama, Toyonaka, Osaka 560-8531, Japan}, F Strigari$^1$, A Hendricks$^1$, D Regesch$^1$, Z Hu$^2$,
Hua Wu$^1$, K-D Tsuei$^{4,5}$, Y F Liao$^4$, H H Hsieh$^6$, H-J Lin$^4$, C T Chen$^4$ and L H Tjeng$^2$}
   \address{$^1$ II. Physikalisches Institut, Universit{\"a}t zu K{\"o}ln, Z{\"u}lpicher Str. 77, 50937 K{\"o}ln, Germany}
    \address{$^2$ Max Planck Institute for Chemical Physics of Solids, N\"othnitzer Str. 40, 01187 Dresden, Germany}
   \address{$^3$ Department of Quantum Matter, ADSM, Hiroshima University, Higashi-Hiroshima 739-8530, Japan}
    \address{$^4$ National Synchrotron Radiation Research Center (NSRRC), 101 Hsin-Ann Road, Hsinchu 30077, Taiwan}
    \address{$^5$ Department of Physics, National Tsing Hua University, Hsinchu 30013, Taiwan}
   \address{$^6$ Chung Cheng Institute of Technology, National Defense University, Taoyuan 335, Taiwan}

\ead{Jonas.Weinen@cpfs.mpg.de}
\begin{abstract}
We report on a comparative study of the valence band electronic
structure of NiO as bulk material and of NiO as impurity in MgO.
From the impurity we have been able to determine reliably the
parameters which describe the local correlations, thereby
establishing the compensated-spin character of the first
ionization state or the state created by hole doping. Using
bulk-sensitive x-ray photoemission we identify pronounced satellite
features in the valence band of bulk NiO which cannot be explained
by single-site many body approaches nor by mean field calculations.
We infer the presence of screening processes involving local
quasi-core states in the valence band and non-local coherent
many body states. These processes are strong and the propagation of
an extra hole in the valence band of NiO will therefore be accompanied
by a range of high energy excitations. This in turn will make the
observation of the dispersion relations in the Ni $3d$ bands difficult,
also because the effective band width is no more than 0.25 eV as estimated
from multi-site calculations.
\end{abstract}
\maketitle

NiO is a benchmark system in solid state physics. It crystallizes
in the NaCl structure, has a partially filled $3d$ shell
(Ni$^{2+}$ $d^8$), and is an antiferromagnetic insulator with a
N\'eel temperature of 523~K \cite{Roth58}. It was pointed out
early on by de Boer and Verwey \cite{deBoer37} that many of the
properties of the $3d$ transition metal compounds do not agree
with the predictions of band theory, e.g., standard band theory
predicts NiO to be metallic. A qualitative explanation was
proposed in terms of the Mott-Hubbard model \cite{Mott49,
Hubbard63} in which the on-site Ni $3d$-$3d$ Coulomb interaction
plays a decisive role.

An early \textit{ab initio} attempt to fix the shortcoming of band
theory was to treat NiO as a Slater insulator in which the
doubling of the unit cell allows for the existence of a gap
\cite{Oguchi83, Terakura84PRL, Terakura84PRB}. However, the
calculated gap of about 0.2~eV \cite{Oguchi83} turned out to be
much too small: A combined photoemission (PES) and
bremsstrahlung-isochromat (BIS) spectroscopy study showed that the
band gap is 4.3~eV \cite{Sawatzky84} and established thereby the
correlated nature of NiO. The inclusion of a
self-interaction-correction (SIC) or Hubbard $U$ term to the
density-functional formalism may provide a justification for the
magnitude of the experimental band gap \cite{Svane90,Anisimov91}.

\begin{figure}
\centering
\includegraphics[width=80mm]{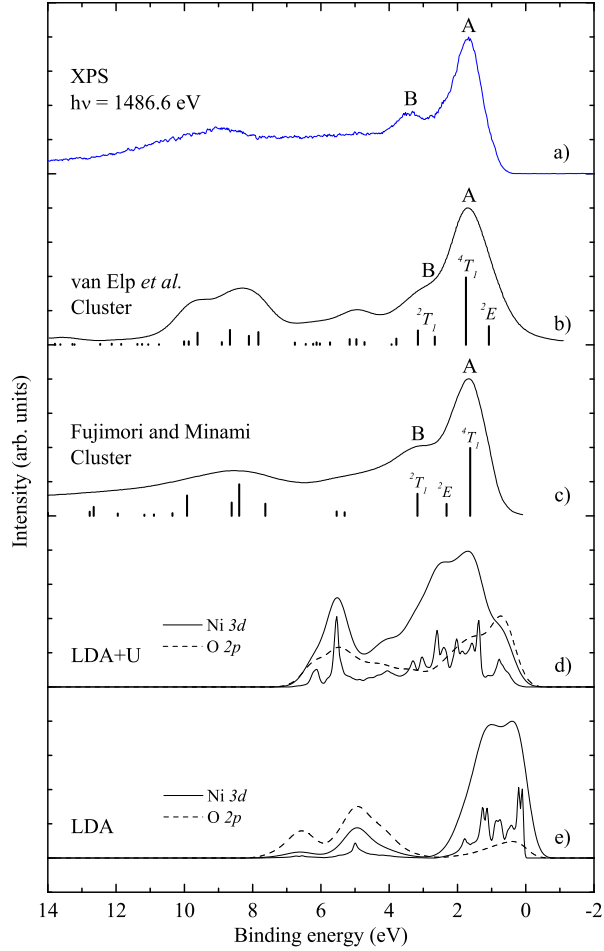}
\caption{(Color online) Valence band {XPS} (1486.6~eV) spectrum of
an \textit{in situ} cleaved NiO single crystal. The results of two
single-site cluster calculations (reproduced from
Refs.~\cite{vElp92} and \cite{Fujimori84}), and LDA and
LDA+U calculations are also included for comparison.}\label{Fig1}
\end{figure}

Yet, one of the most direct methods to critically test the
accuracy of the different approaches, is to determine the
excitation spectrum associated with the introduction of an extra
particle into the system \cite{Almbladh83}. Curve (a) in Fig.~1
displays the valence band x-ray photoemission spectrum ({XPS},
$h\nu$ = 1486.6~eV) of an \textit{in situ} cleaved NiO single
crystal. This spectrum represents essentially the Ni $3d$ spectral
weight since the photoionization cross section of the O $2p$ is
relatively small \cite{Trzhaskovskaya2001}. One can clearly
observe from curve (e) in Fig.~1 that the Ni $3d$ density of
states calculated by band theory (in the local density
approximation, LDA) does not match at all: It has a Fermi cut-off
and the line shape is completely different. The inclusion of the
Hubbard $U$ in the calculations (LDA+U) does not solve the line
shape problem, see curve (d). All this demonstrates the
shortcomings of mean field theories to describe spectra associated
with the fundamental one-particle Green's function of the system
\cite{Anisimov91, Anisimov93}.

A completely different approach is to give up the translational
symmetry of the system in order to focus on the local correlations
and, especially, the dynamics of the propagation of the injected
particle. Curve (c) of Fig.~1 shows the Ni $3d$ spectral weight
from an early cluster configuration-interaction calculation by
Fujimori and Minami \cite{Fujimori84}, which also includes the
full atomic multiplet theory. The agreement with the experimental
spectrum is extremely good. Nevertheless, a later cluster
calculation by van Elp \textit{et al.} \cite{vElp92} arrived at a
less satisfactory result: Peak B has almost disappeared in the
calculation, see curve (b). The prime motivation to use a
different set of model parameters here is to infer that the first
ionization state is low spin ($^{2}E$) \cite{Kuiper89} rather than
the Hund's rule high spin ($^{4}T$), analogous to the case of
Zhang-Rice singlets in the cuprates \cite{ZR88,Eskes88}. Recent
developments combining LDA with dynamical mean field
\cite{Ren06,Kunes07PRL,Kunes07PRB,Yin08} or GW approaches
\cite{Jiang10} yield Ni $3d$ spectral weights which deviate in
important details from the experimental spectrum. These
discrepancies between the experiment and the later theoretical
simulations \cite{vElp92,Ren06,Kunes07PRL,Kunes07PRB,Yin08} do
not provide confidence that one has made progress in understanding
the nature of the first ionization state.

The issues that we need to address now are threefold. First of all
we have to establish whether the {XPS} valence band spectrum in
Fig.~1 is truly representative for bulk NiO. There are reports in
the literature claiming that certain satellite peaks in the Ni
$2p$ spectrum are due to surface effects
\cite{Sangaletti97,Soriano07,Preda08,Mossanek11}. Second, we have
to determine to what extent a single-site many body approach can
be utilized to describe the electronic structure of NiO for which
band formation is also essential. Third, we need to identify the
nature of the first ionization state in the framework of a local
\textit{ansatz}. To this end we measured the valence band of NiO
utilizing the more bulk-sensitive hard x-ray photoelectron
spectroscopy (HAXPES) and we investigated experimentally the
electronic structure of NiO impurities in MgO.


The {XPS} data ($h\nu=1486.6$~eV) on \textit{in situ} cleaved NiO
single crystals were recorded using a Vacuum Generators twin
crystal monochromator Al-$K_{\alpha}$ source and an Scienta
SES-100 analyzer, with an overall energy resolution set to
0.35~eV. The HAXPES data ($h\nu=6500$~eV) were taken at the Taiwan
beamline BL12XU of SPring-8 in Hyogo, Japan using an MBS A-1HE
analyzer. The overall energy resolution was set to 0.35~eV. The
NiO impurity in MgO system was prepared \textit{in situ} as 10-20
nm thin films on polycrystalline Ag by means of molecular beam
epitaxy. The measurements were performed at the 11A1 Dragon
beamline of the NSRRC in Hsinchu, Taiwan. The photoemission
spectra were recorded at the Cooper minimum of Ag $4d$
($h\nu=140$~eV) \cite{Molodtsov00} using a Scienta SES-100
analyzer with an overall energy resolution set at about 0.15~eV.


\begin{figure}
\centering
\includegraphics[width=0.6\columnwidth]{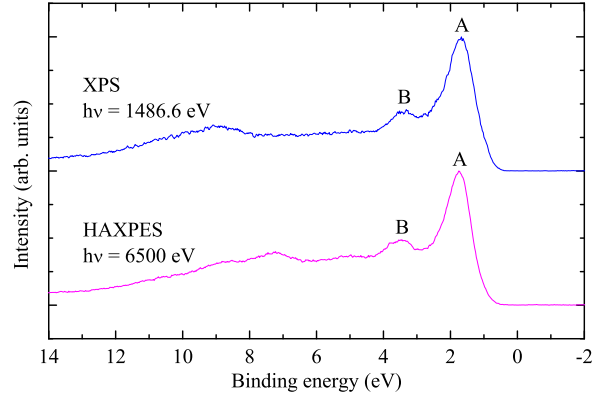}
\caption{(Color online) Valence band photoemission spectra of an
\textit{in situ} cleaved NiO single crystal recorded using
1486.6~eV ({XPS}) and 6500~eV (HAXPES) photons. }\label{Fig2}
\end{figure}

In Fig.~\ref{Fig2}\ we show the valence band photoemission spectra
of a freshly cleaved NiO bulk crystal, taken with a photon energy
of 1486.6~eV ({XPS}) and 6500~eV (HAXPES). By increasing the
photon energy we increase also the kinetic energy of the outgoing
photoelectron and, thus, also the inelastic mean free path. One
can estimate that the probing depth is then enhanced from about
15~\AA\ to roughly 80~\AA\ \cite{Powell00}. We observe that the
spectra are very similar. We, thus, conclude that the {XPS} data
as displayed in Figs.~1 and 2 is representative for the NiO bulk
material and that the contribution of surface effects
\cite{Sangaletti97,Soriano07,Preda08,Mossanek11} can be safely
neglected. To be specific: Peak B is intrinsic for bulk NiO. We
would like to note that increasing the photon energy from
1486.6~eV to 6500~eV does not alter much the Ni $3d$ character of
the spectrum. The O~$2p$ photoionization cross section relative to
that of the Ni $3d$ remains very small, it changes from 1/13 to
only 1/10 \cite{Trzhaskovskaya2001}, meaning that peak B truly
belongs to the Ni $3d$ spectral weight and not to the O $2p$
\cite{Kunes07PRB}.

\begin{figure}
\centering
\includegraphics[width=0.6\columnwidth]{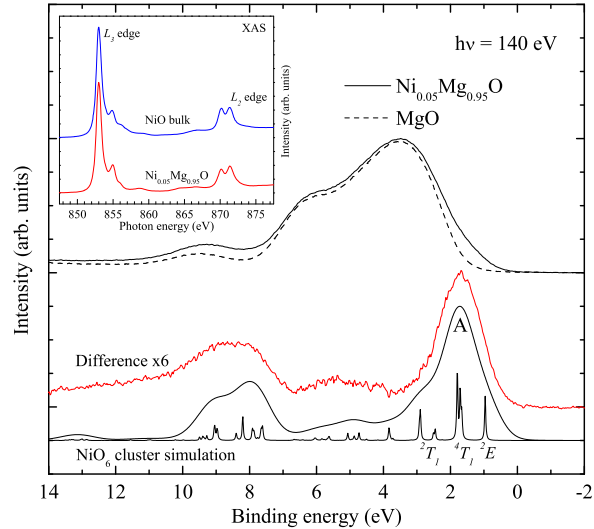}
\caption{(Color online) Extraction of the NiO impurity valence
band photoemission spectrum: Valence band spectra of
Ni$_{0.05}$Mg$_{0.95}$O and an MgO reference, together with the
resulting difference spectrum. Also included is the result of a
single-site NiO$_6$ configuration-interaction cluster calculation.
The inset shows the Ni $L_{2,3}$ x-ray absorption spectrum of the
Ni$_{0.05}$Mg$_{0.95}$O together with that of bulk
NiO.}\label{Fig3}
\end{figure}

The valence band spectrum of the Ni$_{0.05}$Mg$_{0.95}$O impurity
system is shown in Fig.~\ref{Fig3} together with the spectrum of
an MgO reference thin film grown simultaneously under identical
oxygen and substrate conditions. The Ni$_{0.05}$Mg$_{0.95}$O film
(and also the MgO film) was capped by 2 monolayers of MgO in order
to prevent the surface termination to have an effect on the local
electronic structure of the Ni impurity. The inset in the figure
displays the Ni $L_{2,3}$ x-ray absorption spectra of the
Ni$_{0.05}$Mg$_{0.95}$O and the NiO bulk. The spectra are
essentially identical, verifying that the Ni in the
Ni$_{0.05}$Mg$_{0.95}$O has very similar local surrounding
(NiO$_6$ octahedra) as in the bulk.

The valence band spectra of the Ni$_{0.05}$Mg$_{0.95}$O and MgO
systems are normalized to their O $2s$ core level intensities.
Both are dominated by the O $2p$ valence band, yet, there are
clear differences between them due to the presence or absence of
the 5\% NiO impurity. The difference spectrum multiplied by a
factor of 6 is given by the red curve in Fig.~\ref{Fig3}. The line
shape remains the same for films with lower Ni concentrations, as
is the case for that of the Ni $2p$ \cite{Altieri2000}. This curve
represents essentially the Ni $3d$ spectral weight of the NiO
impurity since the photoionization cross section of the Ni $3d$
is an order of magnitude larger than that of the O $2p$ at the
photon energy used \cite{Trzhaskovskaya2001}. Remarkable is that
it is different from the spectrum of bulk NiO as shown in Figs.~1
and 2. The impurity spectrum lacks specifically peak B which is
prominently present in the bulk spectrum.

To interpret and understand the impurity spectrum, we have
performed simulations using the well-proven
configuration-interaction cluster model which includes the full
atomic multiplet theory \cite{Tanaka94,deGroot94,Thole97}. The
simulations have been carried out for a NiO$_6$ cluster using the
program XTLS 8.3 \cite{Tanaka94}.

The bottom curve in Fig.~\ref{Fig3} shows the Ni~$3d$ one-electron
removal spectrum from the cluster calculation. The agreement with
the experiment is very satisfactory. In order to achieve this, we
have started the calculations by using parameter values which were
suggested from earlier studies on NiO
\cite{vElp92,Tanaka94,Alders98,Haverkort04}. We then fine-tune the
parameters describing the octahedral crystal and ligand fields,
and also the difference between the Hubbard $U$ and the
O~$2p$-Ni~$3d$ charge transfer energy \cite{parameters}. The
crucial issue here is to obtain a main line (peak A) without
having another feature appearing at about 2~eV higher energies
(peak B) as was the case in the simulations by Fujimori and Minami
\cite{Fujimori84} and by van Elp \textit{et al.} \cite{vElp92}.
This has implications for the energetics of the states making up
the valence band as we explain in the following.

A detailed look at the cluster calculations displayed in Fig.~1
shows that peak A is given by the $^{4}T_{1}$ final state of the
Ni 3$d^{7}$ multiplet structure while peak B is due to the
$^{2}T_{1}$. Avoiding the appearance of peak B means that the
energy splitting between these two states must be made smaller,
e.g., 1~eV or less. This is what we have done in our simulation in
Fig.~3, using different but equally reasonable parameter values
\cite{parameters}. The consequences for the physics are, yet, quite
far reaching. Given the fact that various x-ray absorption studies
find an effective octahedral crystal and ligand field splitting of
about 1.65~eV \cite{Alders98,Haverkort04}, i.e., the splitting
between the isospin $^{2}T_{1}$ and $^{2}E$ states, we arrive at
the conclusion that the $^{2}E$ must be lower in energy than the
$^{4}T_{1}$ by 0.65~eV or more. This is what we read from our
results in Fig.~3. In other words, our impurity study provides the
spectroscopic evidence that the first ionization state has a
compensated-spin character rather than the Hund's rule high-spin.
This in turn justifies that the ground state of a hole doped NiO
system may indeed be low-spin in nature \cite{Kuiper89}.

\begin{figure}
\centering
\includegraphics[width=0.6\columnwidth]{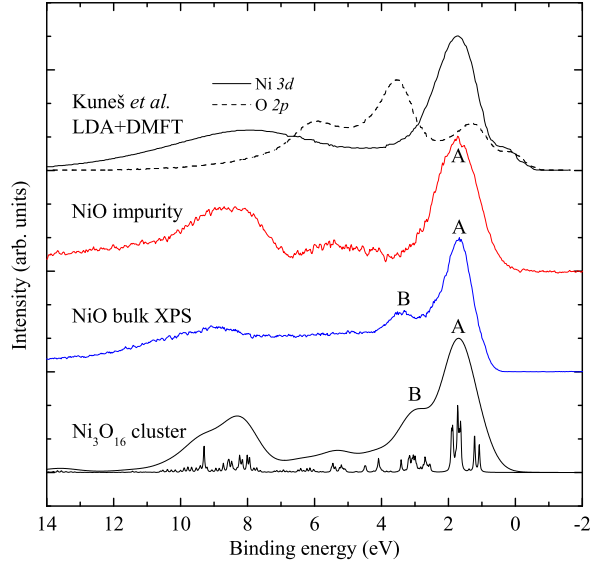}
\caption{(Color online) Comparison of the valence band
photoemission spectra of bulk NiO and NiO impurity in MgO. Also
included is the simulated Ni~$3d$ and O~$2p$ spectral weights of
the NiO valence band from a LDA+DMFT calculation (reproduced from
Ref.~\cite{Kunes07PRB}) and the result of a three-site
Ni$_3$O$_{16}$ configuration-interaction cluster
calculation.}\label{Fig4}
\end{figure}

We now return to the problem of the bulk NiO valence band
spectrum. Fig.~\ref{Fig4} shows the Ni $3d$ spectral weight taken
with {XPS} and compares it with the spectra of the NiO impurity
and of the single-site {LDA}+{DMFT} calculation \cite{Kunes07PRB}.
One can clearly observe that peak B is absent in the impurity as
well as in the Ni $3d$ spectral weight of the single-site
calculation. In fact, one could infer that the calculation
reproduces quite well the impurity spectrum, with perhaps some
discrepancies due to the incomplete implementation of the
multiplet structure of the on-site Coulomb interactions. Yet, the
discrepancy with the bulk spectrum strongly suggests that the
origin of peak B must be sought in non-local correlations, i.e.,
effects which cannot be included in a single-site approach.

Our suggestion is that peak B is due to non-local screening
processes involving the formation of low-energetic coherent many
body states on neighboring NiO clusters, which are of the $^{2}E$
type as we have shown above. The mechanism is analogous as
proposed earlier for the Ni $2p$ core level spectrum of bulk NiO
\cite{vVeenendaal93}, but the application of it for the valence
band is only valid for local states which are relatively stable
against band formation. This may not be applicable for the $^{2}E$
state, which is a state in which a hole is injected in an $e_g$
orbital starting from the $3d^{8}$ $^{3}A_{2}$ ground state
\cite{Haverkort04}. This hole can be expected to readily propagate
in the lattice since the hopping between the Ni $3d$($e_g$) and O
$2p$($\sigma$) orbitals are rather large
\cite{vElp92,vVeenendaal93}, yet, it may leave behind an
energetically costly wake of wrong spins in the antiferromagnetic
lattice. In any case, it would not be meaningful to describe its
band formation as a low-energy screening process involving
neighboring $^{2}E$ states \cite{Taguchi08}.

\begin{figure}
\centering
\includegraphics[width=0.5\columnwidth]{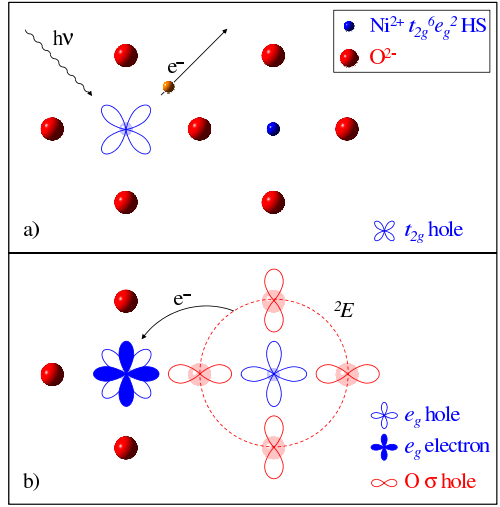}
\caption{(Color online) Non-local screening in valence band
photoemission on NiO: (a) Creation of the atomic-like (quasi-core)
$^{4}T_{1}$ hole state by the photoemission process. (b) Screening
by a next nearest neighbor NiO$_6$ cluster producing a coherent
low-energetic $^{2}E$ hole state there.}\label{Fig5}
\end{figure}

However, for the main peak of the bulk NiO spectrum, i.e., the
$^{4}T_{1}$ state, we infer that we can make a meaningful
approximation by using the coherent $^{2}E$ screening model. The
$^{4}T_{1}$ consists of a hole injected into the $t_{2g}$ orbital,
and its ability to move is rather limited since the overlap
between the Ni $3d$($t_{2g}$) and O $2p$($\pi$) is small. One
could consider the $^{4}T_{1}$ as a localized quasi-core state. We
then can invoke the non-local screening process as follows: After
the creation of the $^{4}T_{1}$ state, an $e_g$ electron from a
neighboring NiO cluster hops onto the Ni site, leaving behind a
coherent $^{2}E$ hole state on that neighbor. A sketch for this
process is given in Fig.~\ref{Fig5}. These two states are
energetically almost degenerate \cite{vVeenendaal93}, and the Ni
$3d$($e_g$) and O $2p$($\sigma$) hybridization between them is
then strong enough to produce two peaks: Not only the main peak A
but also the satellite peak B.

To confirm our assignments, we have performed a Ni$_3$O$_{16}$ cluster calculation consisting of three edge-shared NiO$_6$ octahedra. While all the O~2$p$ and Ni~3$d$ orbitals are included for the NiO$_6$ octahedron in the center where the photo-excitation takes place, those on the other parts of the cluster are replaced by a
reduced basis set using the method in Ref.~\cite{Tanaka99}. The
results are displayed in Fig.~\ref{Fig4}\ and demonstrate the presence of both peaks A and B. Note that we have used the same parameters as for the single-site calculation which produces only peak A, see Fig.~\ref{Fig3}. We also note that the energy difference between peaks A and B is somewhat smaller and the intensity of peak B is slightly larger than those of the experiments. This can be explained by the fact that the number of neighboring Ni sites is only two in the Ni$_3$O$_{16}$ cluster: the energy difference will increase and the intensity of peak B will decrease for a larger number of neighboring sites \cite{Tanaka99}.


To summarize: We have succeeded to determine reliably the Ni $3d$
valence band spectra representative for bulk NiO as well as for
NiO as an impurity system. From the impurity data we are able to
extract the local electronic structure and the correlations
herein, thereby establishing firmly the compensated-spin character
of the first ionization state. By comparing the bulk with the
impurity system, we were able to identify features in the bulk NiO
spectrum which are caused by screening processes involving local
quasi-core valence band states and non-local low-energetic many
body states. These processes are strong and cause considerable
redistribution of the spectral weight when an extra hole is injected
into NiO. This explains why fine features related to the dispersion
of the Ni $3d$ band is difficult to detect \cite{Shen90}, also since the
effective band width is only of order 0.25 eV as estimated from our
multi-site cluster calculations.


We gratefully acknowledge the NSRRC and the SPring-8 staff for
providing us with beamtime. We would like to thank L. Hamdan for
her skillful technical and organizational assistance. The research
in Germany is partially supported by the Deutsche
Forschungsgemeinschaft through SFB 608 and FOR 1346.

\section*{References}


\end{document}